\documentclass{aa}
\usepackage{graphicx}
\usepackage{amsmath}
\usepackage{amssymb}
\usepackage{natbib}
\bibpunct{(}{)}{;}{a}{}{,}
\newcommand{\bm}{\boldmath}
\newcommand{\bv}[1]{\mbox{\bm $#1$}}
\newcommand{\ud}{\mathrm{d}}
\begin{document}
   \title{Quiet time particle acceleration in interplanetary space}


   \author{
           F. Lepreti \inst{1},
           H. Isliker \inst{1},
       K. Petraki \inst{1,2},
       \and L. Vlahos \inst{1}
           }

   \offprints{F. Lepreti}

   \institute{Department of Physics, Aristotle University of Thessaloniki,
              54124 Thessaloniki, Greece\\
              \email{lepreti@astro.auth.gr}
         \and
              Department of Physics and Astronomy, UCLA, Los Angeles,
          California 90095, USA
             }

   \date{Received ........ / Accepted ........}

   \abstract{
   We propose a model for the acceleration of charged particles in
   interplanetary space that appear during quiet time periods, that is,
   not associated with solar activity events like intense flares or
   coronal mass ejections. The interaction
   of charged particles with modeled turbulent electromagnetic fields,
   which mimic the fields observed in the interplanetary medium,
   is studied.
   The turbulence is modeled by means of a dynamical system,
   the Gledzer-Ohkitani-Yamada (GOY) shell model,
   which describes the gross features
   of the Navier-Stokes equations. The GOY model is used to build
   a 3-D velocity field, which in turn is used to numerically solve the
   ideal magneto-hydrodynamic (MHD) induction equation,
   while the electric field is calculated from the ideal Ohm's law.
   Particle acceleration in such an environment is investigated by
   test particle simulations, and the resulting energy distributions
   are discussed and compared to observations of suprathermal electrons
   and ions during quiet periods in interplanetary space.
   }

   \authorrunning{F. Lepreti et al.}

   \maketitle
%

\section{Introduction}

The observations of high energy, suprathermal particle populations
in interplanetary space reveal the presence of a rich variety of
physical characteristics and processes [see \citet{reames} for
a review]. The information recovered from time profiles of
particle fluxes, energy spectra, element abundances, ionization states,
etc., is essential in determining many properties of the sources
and of the mechanisms of particle acceleration. The remarkable
heterogeneity found in the detected energetic particle events
is related, in large part, to the existence of
different sources of acceleration in interplanetary space.
Suprathermal populations can be produced by solar flares,
collisionless shock waves driven by Coronal Mass Ejections (CMEs),
Corotating Interaction Regions (CIRs), or the heliospheric termination
shock.

For many years, the attention of
researchers has been focused on suprathermal particles with energies
above a few hundred keV, due to the fact that the spacecraft instruments
lacked the sensitivity needed
to investigate the energy range from the solar wind thermal
plasma up to a few hundred keV. In recent years, however, with the launch
of the Ulysses, Advanced Composition Explorer (ACE), and WIND spacecrafts,
this gap
has been filled and observations of ``quiet time'' suprathermal particles,
that is, not associated with the abrupt energization events mentioned
above, have become available.

The electron spectrum, measured between
$\sim$5~eV and $\sim$100~keV by the 3-D Plasma and Energetic
Particles Experiment \citep{lin95} on the WIND spacecraft during
a quiet period, has been
investigated by \citet{lin98}. A Maxwellian core
dominates the spectrum from $\sim$5~eV to $\sim$50~eV, while a
hotter population, the so-called solar wind halo \citep{feldman75},
takes over in the range between $\sim$100~eV
to $\sim$1~keV, due to the escape of coronal thermal electrons with temperature
of $\sim 10^6$~K. However, these WIND observations have made possible
the identification of a third, much harder component, which has been denoted
the ``super-halo'', with energies
from $\sim$2~keV up to $\gtrsim$100~keV and an approximate power law
shape with exponent $\sim 2.5$. The angular distribution of these
``super-halo'' electrons is nearly isotropic. According to \citet{lin98},
this high energy population is not solar in origin,
since this would imply a continuous production and escape of
electrons with such energies from the Sun.
It has been suggested \citep{lin98} that the ``super-halo'' tail is due to
acceleration by CIRs beyond 1~AU, but clear evidence for correlations
with CIRs or solar active regions have not yet been found.

The velocity distributions of solar wind ions from 0.6 to 100~keV/e,
measured using the
Solar Wind Ion Composition Spectrometer (SWICS) instruments
on Ulysses and ACE \citep{gloeckler92,gloeckler95}, have been
studied in \citet{gloeckler99}, \citet{gloeckler00}, and
\citet{gloeckler03}. One of the most important findings of these
works is that the speed distributions of H$^+$, He$^+$ and He$^{++}$ ions
show well developed, approximate power law tails during quiet time
periods, that is, far from shocks, CIR compressions and other disturbances. 
This indicates the presence of a population of highly suprathermal ions
at all times. The power law exponents of these tails, which extend
over the whole measurement range, are between $\sim 5$ and $\sim 5.5$
in the slow, in-ecliptic solar wind, and $\sim 8$ in the super-quiet
fast wind coming from polar coronal holes. Comparing ACE observations at
$\sim 1$~AU to Ulysses observations at $\sim 5$~AU, the authors
also found that the tails are continuously regenerated
in the out-flowing solar wind, overcoming the cooling related to
the wind expansion. The main question arising from these observations
is how these ubiquitous suprathermal ions are produced in the quiet
solar wind when there are no shocks, CIRs or other
disturbances observed locally. \citet{leroux01} suggested that
pickup ions might be accelerated by large-scale turbulent electric
fields directed along the background magnetic field. In order to
explore this possibility, they presented a numerical model for
gyrotropic, pitch-angle dependent pickup ion transport between
the Sun and the Earth based on standard kinetic theory for
charged particles. The ion kinetic transport equation used in
their model includes a Gaussian random value
representation of the large-scale field-aligned electric field
fluctuations averaged over the characteristic length and time
scales of MHD turbulence in the low-latitude solar wind. 
The authors choose the standard deviation of these electric fields
by requiring the reproduction of observed accelerated pickup
ion spectra and in this way they were able to obtain a qualitative
agreement of the results of their model with the suprathermal
He$^+$ spectra in the slow low-latitude solar wind, observed at
$\sim 1$~AU.

\citet{kirsch} recently presented an analysis of interplanetary 
suprathermal ions, based on
measurements performed with the SMS experiment \citep{gloeckler95}
on the WIND spacecraft, in the range 0.5-225~keV/e.
The authors investigated particle bursts in which
high energy protons in the range 5-100~keV are observed in association
with distinct decreases of the magnetic field magnitude.
They considered only events not associated with
shocks, CIRs, and magnetospheric disturbances. As a consequence
of the the observed magnetic field behaviour, they suggest that
these bursts could be the result of a local reconnection process, or, 
alternatively, they propose that inductive electric fields
[i.e. $\nabla \times \bv{E} = - (1/c) \partial \bv{B} / \partial t$]
could be a possible explanation for the observed
particle acceleration.

Here, we investigate with a new approach the possibility that the high
energy particles observed in the interplanetary space during quiet time
periods are due to a process of stochastic acceleration in the turbulent
electromagnetic fields present in the heliospheric plasma.
Recently,
the stochastic acceleration process in turbulent electromagnetic fields
has been investigated with numerical experiments in which
test particle simulations are performed in field configurations that are
obtained from the solution of the magneto-hydrodynamic (MHD) equations
\citep{nodes,dmitruk}. The authors suggest that these simulations
could potentially be applied to astrophysical problems.
As an alternative approach, other authors
have studied the motion of test particles in electromagnetic
fields built up by means of suitable models for
particular applications, like the Earth's magnetotail \citep{veltri}
or the solar corona \citep{arzner}.
With respect to other approaches, test particle simulations offer
the possibility to describe some peculiar features of particle
acceleration in turbulent fields, especially the possibility for
particles to be trapped, possibly in or around strong, coherent electric
field regions, leading to effective acceleration even for low initial energies
\citep{ambrosiano,dmitruk}.

In the present paper, we  present a model for stochastic, quiet time
particle acceleration in the interplanetary space, based on turbulent
electromagnetic fields constructed by means of  the so-called
Gledzer-Ohkitani-Yamada (GOY) shell model \citep{gledzer,yamada},
which mimics the nonlinear dynamics of fluid turbulence.
The use of such a simplified description
of turbulence implies that we can describe only some basic features of
the nonlinear interactions occurring in turbulent fluids, with the
advantage that we can simulate turbulent
electromagnetic field configurations without the computational difficulty
of solving the full MHD equations. The magnetic
field is determined through the MHD induction equation, assuming weak
magnetic fields so that the back reaction onto the plasma can be neglected.
The electric field is given by Ohm's law. In this framework, we
perform test particle simulations, concentrating mainly on the energetics
of the injected particles, i.e. electrons and ions.

A basic description of the model is given in Sect. \ref{sec-model}.
In Sect \ref{sec-results}, we present the numerical simulations performed
and the results obtained from them. Discussions and conclusions are
given in Sect. \ref{sec-conclu}.


\section{The model}
\label{sec-model}

The model describes the acceleration of charged test
particles in turbulent electromagnetic fields, as obtained
from a dynamical system model of turbulence. In general, the
macroscopic description of a plasma is given by the
magneto-hydrodynamic (MHD) equations [see e.g. \citet{boyd}].
Here, we consider particle acceleration events
occurring in regions of the interplanetary space
where the magnetic field is weak, so that we can
restrict ourselves to weakly magnetized plasmas, which implies that the
temporal evolution of the plasma is governed by the velocity field.
In order to build up the velocity field configurations,
we use the so-called GOY shell model, through which it is possible
to generate a turbulent, incompressible
3-D velocity field $\bv{V}(\bv{r},t)$, as will be explained in Sect.
\ref{sec-vfield}.

The velocity field is used to obtain numerical solutions
of the MHD induction equation for a perfectly
conducting plasma, namely
\begin{equation}
{\partial \bv{B} \over \partial t} =
\nabla \times ( \bv{V} \times \bv{B} ) \; ,
\label{eq-B-id}
\end{equation}
where $\bv{B}$ is the magnetic field. The dissipative term
$\mu \nabla^2 \bv{B}$ (where $\mu$ is the magnetic diffusivity)
is not taken into account, since the
plasma in the interplanetary space can be considered collisionless
to a good approximation [see e.g. \citet{montgomery}]. The electric field
is then computed from the ideal Ohm's law
\begin{equation}
\bv{E} = -{1 \over c} \bv{V} \times \bv{B} \; ,
\label{eq-E-id}
\end{equation}
where the resistive term $\eta \bv{j}$ (with $\eta$ being the resistivity
and $\bv{j}$ the current density, respectively) is again neglected.
As already mentioned before, the feedback of the magnetic
field on the velocity field (that is, the effect of the Laplace force) is
not considered, as the velocity field is given independently and the
MHD momentum equation is not considered. In other words, we suppose
that the magnetic energy density is much smaller than the kinetic energy
density of the turbulent flow.

To investigate the acceleration of test particles in the
electromagnetic fields generated as described above, we consider the
relativistic equations of motions of a charged particle in an external
electromagnetic field:
\begin{equation}
\frac{\ud \bv{r}}{\ud t} = \bv{v} \\
\label{acc1}
\end{equation}
\begin{equation}
\frac{\ud \bv{p}}{\ud t} =
 q \bv{E} + \frac{q}{c} \bv{v} \times \bv{B} ,
\label{acc2}
\end{equation}
where $\bv{r}$, $\bv{v}$, and  $\bv{p}$ are the position, velocity,
and momentum of the particle, respectively,
$c$ the speed of light, and $q$ the charge of the particle.
For the numerical integration, we express Eq. (\ref{acc2})
in terms of velocity ($m \gamma \bv{v}= \bv{p}$),
\begin{equation}
\frac{\ud \bv{v}}{\ud t} =
 \frac{q}{\gamma m} \bv{E} +
\frac{q}{\gamma m c} \bv{v} \times \bv{B}
-\frac{q}{\gamma m c^2} \bv{v} (\bv{v} \cdot \bv{E}) ,
\label{acc3}
\end{equation}
where $\gamma=1/\sqrt{1-v^2/c^2}$, and $m$ is the mass of the particle.


\subsection{Construction of the 3-D fluid velocity field}
\label{sec-vfield}
The 3-D, time-dependent velocity field $\bv{V}(\bv{r},t)$
is constructed by means of the so-called GOY shell
model for turbulence. Shell models [see \citet{DynSys} for a complete
review] are dynamical systems designed to represent in a
simplified way the spectral form of the equations which describe
turbulent fluids. They were originally proposed by
\citet{obukhov}, \citet{novikov}, and \citet{gledzer} for hydrodynamic
turbulence. The GOY shell model \citep{gledzer,yamada} has been extensively
investigated, both analitically and numerically 
\citep{yamada, jensen, biferale}. In the following, we describe some
basic characteristics of the GOY model.

The main idea of the GOY shell model is to mimic the Navier-Stokes
equations by a dynamical
system in which the velocity field fluctuations at different
length scales are represented by scalar variables $u_n(t)$.
To this aim, the Fourier space is divided into N shells, with the
associated wave number denoted by $k_n$, where the shell index $n$ is
discrete. The scalar, complex
variable $u_n(t)$ is associated with the $n$-th shell, and the nonlinear
dynamics of turbulent fluids are modeled by quadratic couplings
among nearest and next nearest neighbour shells, following the assumption
that the nonlinear interactions are local in the $k$ space. The coefficients
of the nonlinear terms are determined by imposing the conservation of the ideal
invariants of the Navier-Stokes equations.   The equations of evolution of the
dynamical variables $u_n(t)$ are \citep{DynSys}
\begin{eqnarray}
& & \left(\frac{\ud}{\ud t}+ \nu k_n^2 \right )u_n = \nonumber \\
& & i(\alpha k_{n+1}u_{n+2}^* u_{n+1}^* +
\beta k_{n}u_{n+1}^* u_{n-1}^* + \nonumber \\
& & \gamma k_{n-1}u_{n-1}^* u_{n-2}^*) + \delta_{mn}f_n
\; ,
\label{goy}
\end{eqnarray}
\begin{small} \begin{flushright} where $n = 1, ... ,N$.
\end{flushright} \end{small}
The parameter $\nu$ stands for the kinematic viscosity, while 
$\delta_{mn}f_n$ is
a stochastic forcing term acting on the shell $m$, one of the first shells, 
providing a
constant average energy flux into the system 
($\delta_{mn}$ is the Kronecker symbol). The wave numbers
are chosen to follow the relation
\begin{equation}\label{k}
 k_n = k_{\mathrm{0}} h^n \; ,
\end{equation}
where $k_{\mathrm{0}}$ and $h$ are constant ($h>1$, usually $h=2$),
and $n>1$.
The shells are thus equally spaced in a logarithmic scale,
which is justified by the fact that in fully developed turbulence the energy
spectrum  in the nonlinear, inertial range follows a power law.

One of the main advantages of shell models over numerical
simulations of the Navier-Stokes equations is that they can be
investigated at much higher Reynolds numbers. They provide a good
description of the scaling properties of fully developed turbulence
in the inertial range, even if, being
scalar models, they do not include information about the spatial structures
of turbulence.
From the scalar variables $u_n(t)$ we can generate 
an incompressible velocity
field $\bv{V}(\bv{r},t)$ in the real 3-D space by applying a simple numerical
algorithm \citep{DynSys}.
The dynamical variables $u_n(t)$ of the shell model are supposed to
represent the coefficients
of a Fourier expansion with wave vectors $\bv{k}$ in a shell of
radius $|\bv{k}| = k_n$.
We introduce a set of vectors $\bv{k_n}$:
\begin{equation}
\label{k-vector} \bv{k_n} = k_n\bv{e_n} \; ,
\end{equation}
where $\bv{e_n} = \{e_n^{(1)},e_n^{(2)},e_n^{(3)}\}$ are randomly
chosen vectors of unit norm. The components
$V_j(\bv{x},t)$, $j = 1,2,3,$ of the velocity field are obtained by the
analogue to an inverse Fourier transform,
\begin{equation}
V_j(\bv{r},t) = \sum_{n=1}^N
C_n^{(j)}[u_n(t) e ^{i\bv{k_n \cdot r}} + c.c.] \; ,
\label{eq-vgoy}
\end{equation}
where the coefficients $C_n^{(j)}$ are of order $O(1)$.

In order to satisfy the incompressibility constraint
$\nabla \cdot \bv{V} = 0$, the  vectors
$\bv{e_n} = \{e_n^{(1)},e_n^{(2)},e_n^{(3)}\}$ and the coefficients
$C_n^{(j)}$ must satisfy the condition
\begin{equation}
\sum_{j=1}^{3} C_n^{(j)} e_n^{(j)} = 0 \; , \; \; \; \; \; \forall n \; .
\end{equation}

\section{Numerical simulations and results}
\label{sec-results}
\begin{figure*}
\centering
\includegraphics[width=17cm]{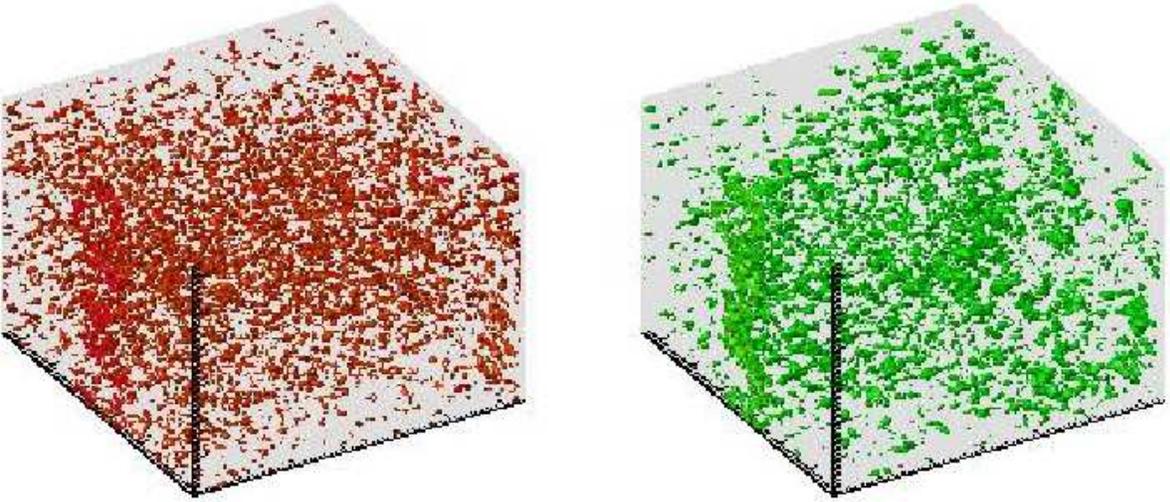}
\caption{3-D visualization of the magnetic field intensity $|\bv{B}|$ above
the threshold $B_{th} = 1.2 \times 10^{-4}$~G (left panel) and of the
electric field intensity $|\bv{E}|$ above the threshold
$E_{th} = 1.2 \times 10^{-7}$~statvolt~cm$^{-1}$ (right panel).}
\label{fig-bfef}
\end{figure*}
The numerical investigation consists of two main steps: 1) the calculation
of the magnetic and electric fields from Eqs. (\ref{eq-B-id}) and
(\ref{eq-E-id}), in which the velocity field given by Eq. (\ref{eq-vgoy})
is used; 2) test particle simulations through the numerical
solutions of Eqs. (\ref{acc1}) and (\ref{acc3}).

\subsection{Magnetic and electric field calculation}
The GOY shell model equations Eq. (\ref{goy}) are solved by using a 4-th order
Runge-Kutta integration algorithm, using $N=22$ shells. The kinematic
viscosity in the shell model is assumed to be $\nu = 10^{-7}$.
Once the shell model has reached a statistically stationary state,
we start the numerical integration of the MHD induction equation
[Eq. (\ref{eq-B-id})], by using
the Wilson upwind scheme \citep{wilson},
imposing free outflow boundary conditions and
using the velocity field $\bv{V}(\bv{r},t)$ from Eq. (\ref{eq-vgoy}).
The initial magnetic field $\bv{B_0}(\bv{r})$
is given by a random perturbation constructed through a sum
of Fourier modes
$\bv{B_0}(\bv{r}) = \sum_{\bv{k}} \bv{B_0}(\bv{k})
\cos (\bv{k} \cdot \bv{r} + \varphi_{\bv{k}})$,
with Gaussian distributed amplitudes $|\bv{B_0}(\bv{k})|$,
random phases $\varphi_{\bv{k}}$ and with the constraint
$\bv{k} \cdot \bv{B_0}(\bv{k}) = 0$ imposed, as it follows
from $\nabla \cdot \bv{B} = 0$. The size of the grid is $64^3$.
The evolution of the system is followed over a time interval
$2 \tau_e$, where $\tau_e$ is a typical eddy turnover time,
given by $\tau_e = L / V_{\mathrm{rms}}$ , $L$ being the size of the simulation
domain and $V_{\mathrm{rms}}$ the rms velocity. The condition that the kinetic
energy density is much smaller than the magnetic energy density is
checked during the time evolution. We also monitor the value of
$\nabla \cdot \bv{B}$, evaluated through a standard finite
differences scheme, and verify that it does not vary
significantly with respect to the initial value, which can
be considered zero within the numerical error.

The equations are solved in non-dimensional form, and suitable
rescaling factors are applied to describe quiet time periods in
interplanetary space. The applied forcing term leads to
an rms velocity field intensity of
$V_{\mathrm{rms}} \simeq 3.2 \times 10^7$~cm~s$^{-1}$.
This value is slightly larger that the rms velocity estimates
obtained from a 30 year dataset of solar wind observations,
which gave values around $7 \times 10^6$~cm~s$^{-1}$ \citep{breech}.
This means that in our simulations we are assuming a slightly enhanced
turbulence level with respect to the average.
The linear size of the simulation box is assumed to be
$L = 2.2 \times 10^{10}$~cm.  The rms value of the magnetic field intensity,
after the rescaling, is $B_{\mathrm{rms}} \simeq 7.4 \times 10^{-5}$~G,
while the rescaled rms value of the electric field intensity is
$E_{\mathrm{rms}} \simeq 5.7 \times 10^{-8}$~statvolt~cm$^{-1}$.
These values are slightly larger than the rms values obtained
from long time datasets of solar wind observations.

To illustrate the 3-D structure of the magnetic and electric
fields, in Fig. \ref{fig-bfef} we present a 3-D visualization of the
regions where the magnetic field and electric field intensities exceed
the thresholds $B_{th} = 1.2 \times 10^{-4}$~G and
$E_{th} = 1.2 \times 10^{-7}$~statvolt~cm$^{-1}$, respectively.

\begin{figure}
\resizebox{\hsize}{!}{\includegraphics{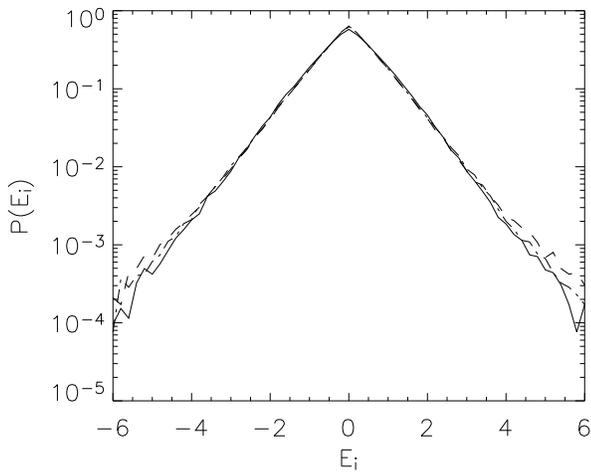}}
\caption{PDFs of the three components of the electric field: the solid,
dashed, and dot-dashed lines represent the PDFs of $E_x$, $E_y$,
and $E_z$ respectively.}
\label{fig-pdf-ef}
\end{figure}

In Fig. \ref{fig-pdf-ef} we present the Probability Density Functions (PDFs)
of the three electric field components, collected at a fixed time from the
entire simulation box.
%
%
In order to compare the different PDFs,
the variables are first translated to zero mean and then normalized to
their standard deviation,
so that all the PDFs have zero mean and unit standard deviation. As it can
be seen, these PDFs are not Gaussian, they exhibit clear exponential
tails. This result is in qualitative agreement with the one-point PDFs
of the observed interplanetary induced electric fields (IEF) presented  in
\citet{breech}. In a more recent work \citep{sorriso}, it has been
shown that the statistical properties of the IEF depend on the wind
velocity, an effect that we do not model here
(see discussion in Sect. \ref{sec-conclu}).

\subsection{Test particle simulations}
In this subsection, we present some results obtained from test particle
simulations of electrons and ions in the electromagnetic field
configurations generated as described in the previous subsection.
Specifically, the magnetic and electric field configurations obtained at
the time $2\tau_e$ during the evolution of Eq. (\ref{eq-B-id}) are used.
The particle motion equations (\ref{acc1}) and (\ref{acc3}) are solved with
a 4th order Runge-Kutta, adaptive step-size scheme. The magnetic and
electric field configurations are kept constant during the time we monitor
the particles, assuming a much slower evolution time for the fluid
velocity field than for the test particles.
Since the magnetic and the electric
fields are given only at a discrete set of points, both fields
are interpolated with a local, 3-D linear interpolation to provide the field
values in between grid-points, wherever they are needed for the integration
scheme. The initial time step used for the integration is set to $0.1 t_g$,
where $t_g$ is the gyration period of the particles at the starting point.

The charged test particles are injected at random
positions within the simulation box, with velocities extracted from 
the tail of a
Maxwellian distribution with temperature $T = 1.15 \times 10^5$~K
(corresponding to $\sim 10$~eV), which represents
a typical value in the interplanetary space at 1~AU from the Sun.
The threshold velocity used to select only the particles in the tail
of the Maxwellian is $2 v_{th}$, where $ v_{th}$ is the thermal
velocity. The choice to use only the tail of the initial Maxwellian
distribution is due to the fact that we do not model the collisional 
processes occurring in the interplanetary space plasma, so we assume
that only the high energy part of the distribution participates
in the acceleration process.
If a particle leaves the simulation box before the end of its
tracing time interval, it is reinjected into the box at a random point
on the surface opposite to the one through which it had left.
The maximum tracing time interval used in this work is $t = 300$~s,
which is much smaller
than the typical collision times for electrons and ions with velocities 
larger than twice the thermal velocity [see e.g. \citet{montgomery}], 
so that we can neglect collisional effects.

\begin{figure}
\resizebox{\hsize}{!}{\includegraphics{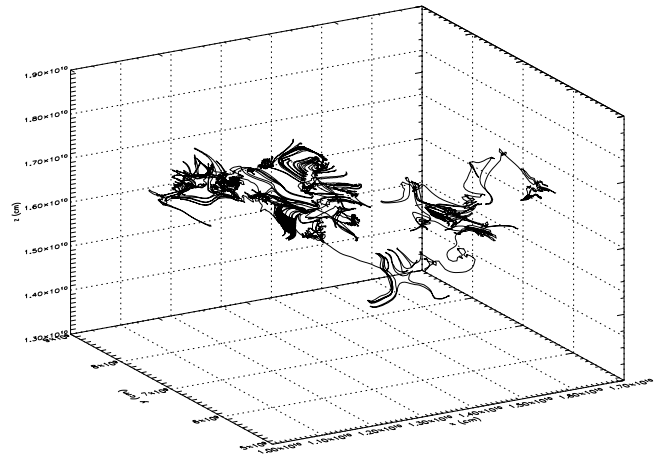}}
\caption{Trajectory of a test electron.}
\label{fig-eltraj}
\end{figure}

\begin{figure}
\resizebox{\hsize}{!}{\includegraphics{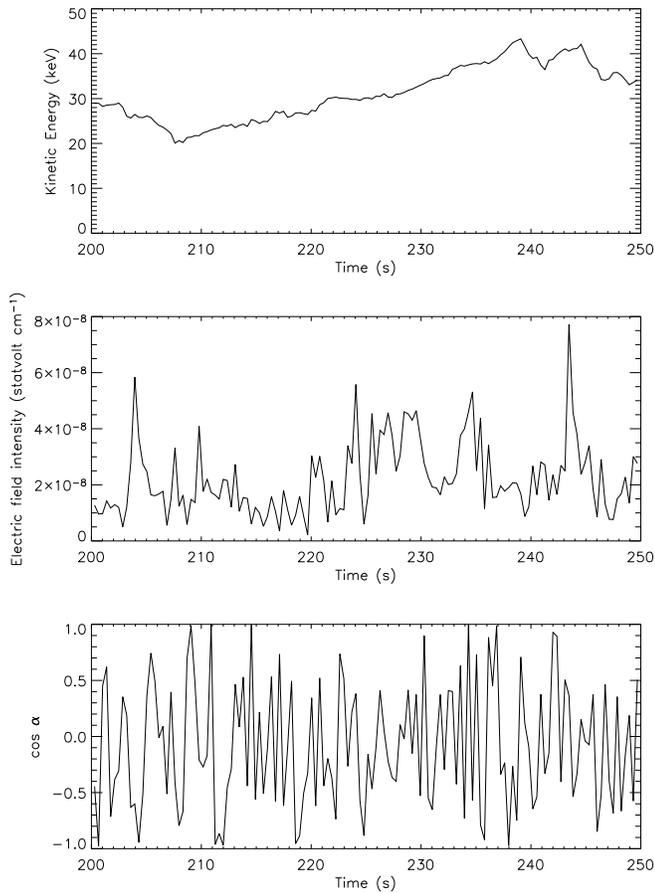}}
\caption{Part of the time evolution of the kinetic energy of a test electron
(upper panel),
the associated electric field intensity along the trajectory (middle panel),
and the cosine of the angle $\alpha$ between the particle velocity and the
electric field (lower panel).}
\label{fig-el-ee}
\end{figure}

In Fig. \ref{fig-eltraj}, a sample trajectory of a typical test electron is
shown.
 It can be seen that, during its irregular motion, the particle visits
both regions where it remains trapped for some time and regions where its
motion exhibits long ``jumps'', as could be expected in a turbulent
field environment.
In Fig. \ref{fig-el-ee}, we report part of the time evolution of the
kinetic energy
of a typical test electron (upper panel), the associated electric field
intensity along the trajectory (middle panel), and the cosine of the angle
$\alpha$ between the particle velocity and the electric field (lower panel).

\begin{figure}
\resizebox{\hsize}{!}{\includegraphics{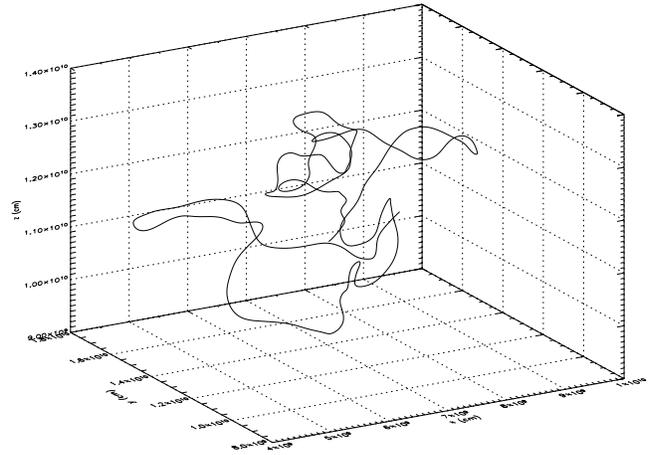}}
\caption{Trajectory of a test proton.}
\label{fig-prtraj}
\end{figure}

\begin{figure}
\resizebox{\hsize}{!}{\includegraphics{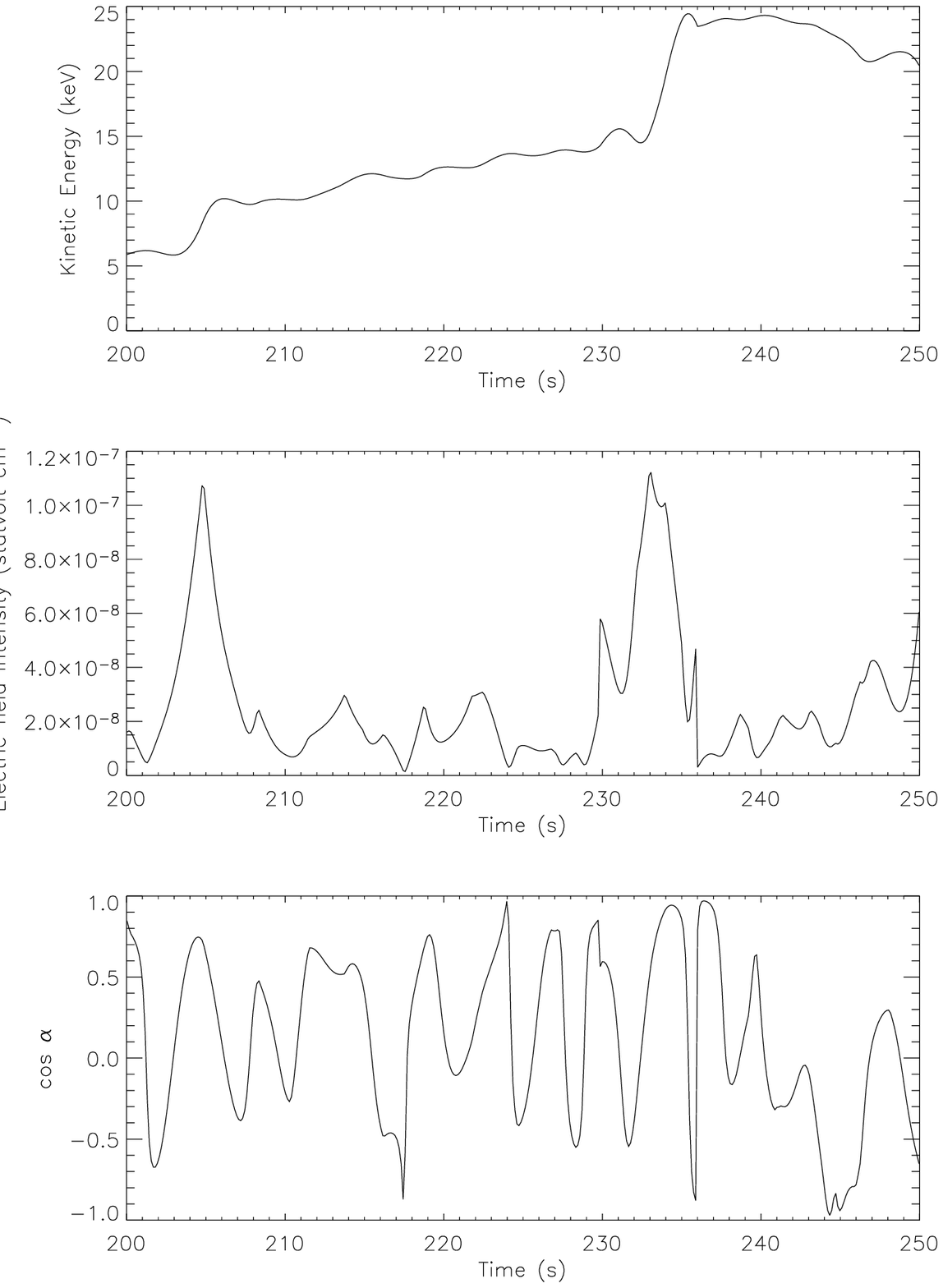}}
\caption{Part of the time evolution of the kinetic energy of a test proton
(upper panel),
the associated electric field intensity along the trajectory (middle panel),
and the cosine of the angle $\alpha$  between the particle velocity and the
electric field (lower panel).}
\label{fig-pr-ee}
\end{figure}

In Fig. \ref{fig-prtraj}, a sample trajectory of a test proton is
shown.
As is expected, the typical shape of the proton trajectories is
substantially different from the one of the electrons, due to the much
larger mass of the protons.
In Fig. \ref{fig-pr-ee}, we report part of the time evolution of the
kinetic energy
of a test proton (upper panel), the associated electric field intensity
along the trajectory (middle panel), and the cosine of the angle $\alpha$
between the particle velocity and the electric field (lower panel).
From this figure, it is more clear, with respect to the figure referring
to electrons (Fig.\ \ref{fig-el-ee}), that the largest kinetic energy 
variations are associated
both with intense electric field spikes and with extended regions of almost
coalignment between particle velocity and electric field.

\begin{figure}
\resizebox{\hsize}{!}{\includegraphics{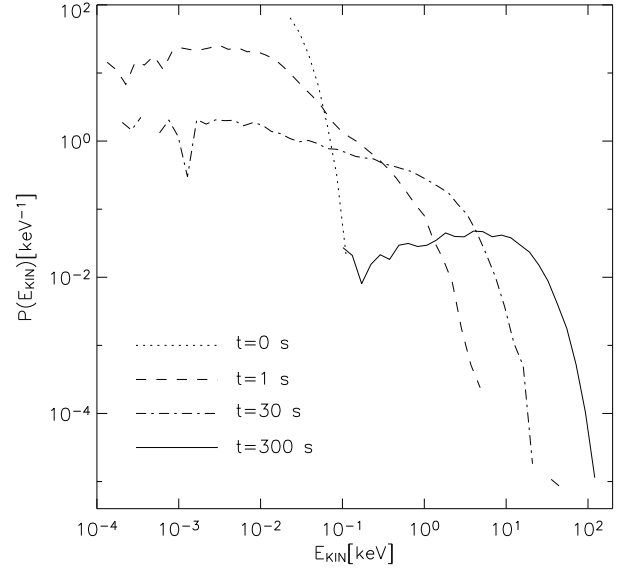}}
\caption{Kinetic energy probability density function of $10^4$ test electrons at the
initial time $t=0$ (dotted curve), and after $t=1$~s (dashed curve),
$t=30$~s (dot-dashed curve), $t=300$~s (solid curve).}
\label{fig-pe-el}
\end{figure}

In Fig. \ref{fig-pe-el}, we report the kinetic energy PDFs of $10^4$ 
test electrons at
the initial time $t=0$~s, and at three successive times $t=1$~s,
$t=30$~s, and $t=300$~s respectively.
At $t=1$~s, the evolution of the initial Maxwellian distribution produces
a tail, extending to $\sim 5$~keV, which does not show a clear, unique 
power law
form, but is more of a double power-law shape. As the time increases, 
the high energy tail tends
to become exponential with maximum energy up to $\sim 100$~keV

\begin{figure}
\resizebox{\hsize}{!}{\includegraphics{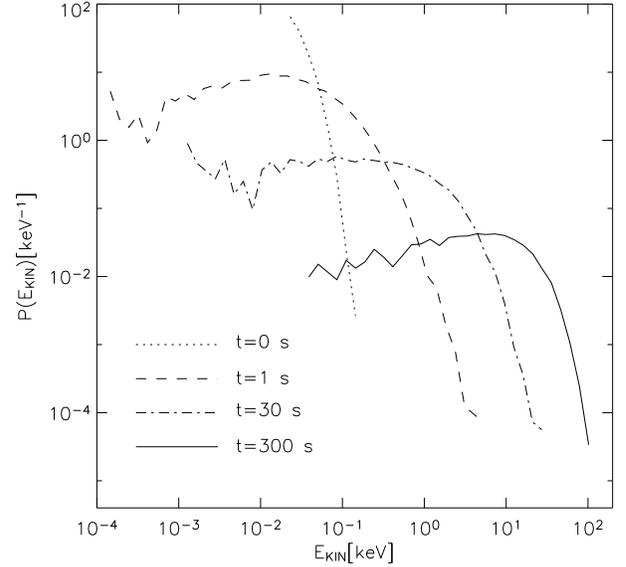}}
\caption{Kinetic energy probability density function of $10^4$ test 
protons at the
initial time $t=0$ (dotted curve), and after $t=1$~s (dashed curve),
$t=30$~s (dot-dashed curve), $t=300$~s (solid curve). }
\label{fig-pe-pr}
\end{figure}

In Fig. \ref{fig-pe-pr}, we report the kinetic energy PDFs of $10^4$ 
test protons at the same times as for the electrons.
The evolution of the proton distributions is qualitatively similar to
the electrons', although the PDF tail at $t=1$~s now displays a reasonably
clear power law scaling that persists until $30$~s. 
However, for $t=300$~s the tail
is narrow and steep, reminiscent of both a steep power-law and an exponential 
distribution, with maximum energy up to $\sim 100$~keV.

\begin{figure}
\resizebox{\hsize}{!}{\includegraphics{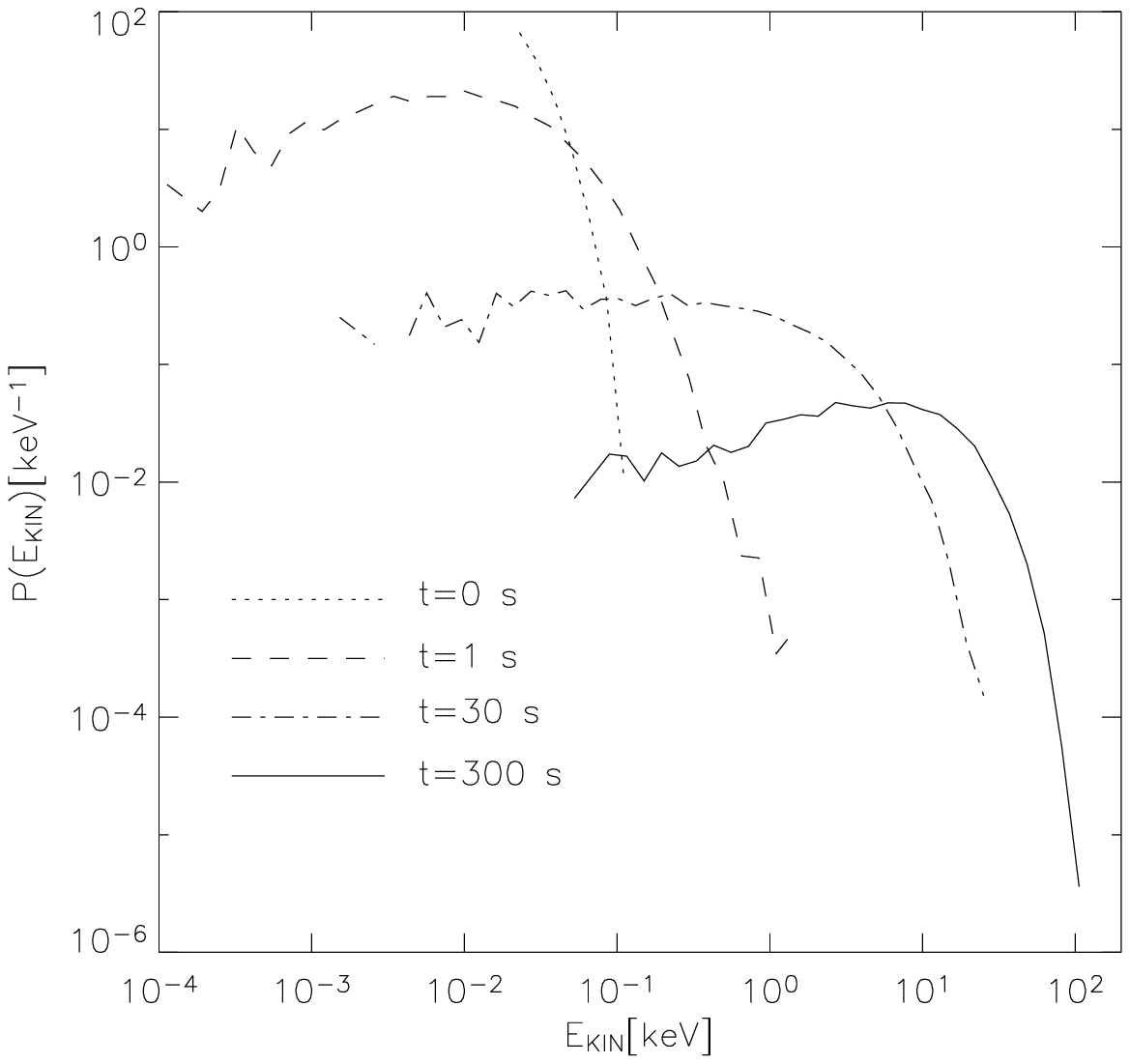}}
\caption{Kinetic energy probability density function of $10^4$ test 
He$^+$ ions
at the initial time $t=0$ (dotted curve), and after $t=1$~s (dashed curve),
$t=30$~s (dot-dashed curve), $t=300$~s (solid curve).}
\label{fig-pe-hep}
\end{figure}

Since observations are also available for other ions, and in particular
for He ions, we also investigated the kinetic energy PDF of He$^+$ ions.
The time evolution of this PDF, shown in Fig. \ref{fig-pe-hep}, is
quite similar to the proton case. For $t=1$~s, the tail of the PDF
extends to lower energies than for the protons, 
again being a power-law shape that persists until $t=30$~s.
For $t=300$~s, energies up to $\sim 100$~keV are  reached, with 
a narrow and steep tail that is difficult to classify.

\section{Discussion and conclusion}
\label{sec-conclu}

In this paper, we have presented a model for turbulent
particle acceleration based on a dynamical system description of
turbulence. The aim of the model is to describe long lasting
particle acceleration processes occurring in the turbulent
interplanetary space during quiet time periods, 
which lead to the appearance of suprathermal
tails at all times in the energy distributions of electrons and
ions (extending up to $\sim$ 100~keV).

The acceleration process has been investigated
by performing test particle numerical simulations in the electromagnetic
fields obtained by numerically solving the ideal MHD induction equation,
which is driven by the velocity field that is calculated using a
dynamical system model
(the so-called GOY shell model) of turbulence. This approach implies that the
magnetic fields are supposed to be weak, or that the magnetic
energy density is much smaller than the kinetic energy density of the flow.

The PDFs of the electric field components have been shown to be 
non-Gaussian, exhibiting exponential tails.
The presence of roughly exponential tails
in the one-point PDFs of the interplanetary induced electric fields (IEF)
has been
recently shown based on data analysis performed on 30 years of
measurements that were acquired by different spacecrafts \citep{breech}.
However, more recently \citep{sorriso}, it has been shown that using
homogeneous datasets with respect to wind velocity
and solar activity, that is, considering short datasets and separating the
data according to slow and fast wind streams,
the exponential tails are recovered
only in the radial components of the electric field, that is, the component
along the Sun-Earth direction, which coincides with the mean wind
direction. Our model is obviously not able to reproduce in detail these
statistical properties of the observed IEF, since we do not take into account
some basic features of the solar wind structure, e.g. its mean bulk velocity,
mean magnetic field structure, etc. In other words, we consider only the
effect of field fluctuations related to the presence of turbulence.

With our approach, we have been able to obtain basic physical insights
into the process of particle acceleration due to 
turbulent electric fields in weakly magnetized plasmas, and to investigate 
the possibility
that this mechanism plays a role in the acceleration of charged particles
to suprathermal energies in interplanetary space during quiet periods.
The trajectories obtained from the simulations indicate that the particles
alternately visit regions in which they are trapped for some time, due
to the effects of turbulence, and other regions where long ``jumps'' are
observed. The observed motion of the test particles suggests that a
Fokker-Planck description of the diffusion process underlying the particle
acceleration is not sufficient to achieve a complete characterization
of the problem. This is one
of the main reasons why a test particle approach is adequate in studying
such situations.

We found that, starting from an initial thermal population, both in the case
of electrons and ions, the initial Maxwellian energy distributions evolve
in time, giving rise to
power-law tails for shorter times, which become very steep and narrow
for the longest time we monitor the particles, so that it is 
difficult to discriminate between exponential and power-law distributions.
At the maximum time we allow, the particles reach 
energies up to $\sim$ 100~keV. The fact that our model is able to reproduce
the observed energies suggests that
a stochastic acceleration mechanism resulting from the turbulence 
that is developed
in the interplanetary space can be at the origin of the ubiquitous
suprathermal tails observed during quiet time periods, even if 
the detailed shape of the observed distributions,
which exhibit approximate power law tails, is not very well
recovered by our model.


The difficulty in reproducing the observations in detail at this stage
is related to some limitations of our model in its current form.
Our main simplifying assumptions are: 
(1) We have assumed
that the magnetic energy density is much smaller than the kinetic
energy density of the fluid. This assumption is often not
fulfilled in the interplanetary space, due to the presence of a
strong Alfvenic component in the solar wind turbulence [see e.g.
\citet{goldstein95,tu95}]. (2)  The detailed properties of
intermittency in solar wind turbulence 
[see e.g. \citet{burlaga91,burlaga92, marsch93, carbone95}] 
are not included in the
model, and this could have an effect especially on the high
energy part of the kinetic energy PDFs. (3) The 
resistivity, which modifies the small-scale structure of the
electric field, is treated as a constant, not taking its possible
dynamic evolution into account [see \citet{dmitruk,arzner}]. 
(4) The large scale 
magnetic field structure, which was not included in the present study, 
but which nevertheless would be interesting to be investigated, will have
a less deciding influence on the energetics.

In order to
overcome the limitation (1), a dynamical system for
MHD turbulence must be considered, instead of a hydrodynamical
model, whereas for the limitation (2), a more appropriate description 
of intermittency should be introduced in the reconstruction
of the 3-D fields from the 1-D scalar fields that are yielded by 
the shell models.
The presented results suggest that on modifying our model in this way,
better compatibility with the observations can be reached. 

\begin{acknowledgements}
This work was partially supported by the Research Training Network (RTN)
``Theory, Observation, and Simulation of Turbulence in Space Plasmas'',
funded by the European Commission (contract No. HPRN-CT-2001-00310).
\end{acknowledgements}

\end{document}